\documentclass[twocolumn,showpacs,aps,prc]{revtex4}

\usepackage[english]{babel} 	
\usepackage{graphics} 
\usepackage[pdftex]{graphicx}
\usepackage{amssymb} 			
\usepackage{amsmath} 			
\usepackage{booktabs}
\usepackage{tabularx}
\usepackage{rotating}

\usepackage{float}
\usepackage{url}	
\usepackage{bm}

\allowdisplaybreaks

\begin{document}

\title{Beta-delayed particle emission and collective rotations}

\author{K. Riisager$^{1}$, E.A.M. Jensen$^{1}$, A.S. Jensen$^{1}$}

\affiliation{$^{1}$Department of Physics and Astronomy, Aarhus University, DK-8000 Aarhus C, Denmark}


\begin{abstract}
  Beta-delayed proton emission in the lower half of the sd-shell will
  involve deformed nuclei. We derive the normalized matrix element
  connecting emission of one particle from an initial rotational
  nuclear state to another final rotating state, and we extract
  selection rules involving the $K$ quantum number. The initial state
  is approximated as having a core identical to the final nuclear
  state. The formalism is then directly applicable to
  $\beta^+$-delayed proton decays of even-$Z$, odd-$N$ nuclei or
  $\beta^-$-delayed neutron decays of odd-$Z$, even $N$ nuclei.  These
  beta-decay results are compared to the outcomes of possible transfer
  reactions.  As an example the beta-delayed proton emission of $^{21}$Mg
  is considered, where new quantum numbers can be assigned to several
  states in $^{21}$Na.
\end{abstract}

\maketitle

\section{Introduction}                            

This paper was motivated by a study of beta-delayed particle emission
in the lower sd-shell, a region where nuclei are known to be well
deformed.  We shall focus mainly on the structure information that may
arise from the particle emission process in the case of deformed
nuclei that exhibit rotational behaviour.  Much information can of
course be extracted by populating the states in particle elastic
scattering or transfer reactions, including detailed information on
the composition of the wave-functions, see e.g.\ chapter 5.3a in \cite{boh69},
but the selection rules of the beta
decay process will give a different feeding pattern, and focussing on
the particle emission rather than state population can give
complementary results. The beta-delayed particle process will give
information on states placed above particle thresholds where emission
of gamma-rays is often negligible.

We shall first derive general expressions for transition matrix
elements, when the particle emitting initial state and the final state
are similar rotations differing only by the extra emitted
particle. They shall be derived based on the general formalism given
in \cite{vil66,sie87}.  The results are then applied to beta-delayed
proton decays of $^{21}$Mg. A more general discussion of the
applicability of the results will also be given at the end of the
paper. Before turning to the concrete formulation of the particle
emission, we recall briefly some relevant aspects of the beta-decay
process.

Beta-decay can give precise information on nuclear structure.
The relevant interaction is extremely weak compared to the
all-decisive, structure determining, strong interaction and the
selection rules of the decay highlight a clean set of nuclear levels.
As we move away from beta-stability, beta-delayed emission of
particles like neutrons, protons or $\alpha$-particles becomes more
probable.
These beta-delayed particles carry information on many aspects of the nuclear
structure of the states in the decay cascade \cite{Poe96,Pfu12}.

The spin-isospin selection rules for allowed beta decay are as
follows. For Fermi decays neither angular momentum $J$, nor isospin $T$, can change, for
Gamow-Teller decays the changes, $\Delta J$ and $\Delta T$, can be $0, \pm 1$ with
no change in parity. All $\beta^-$ decays (except that of the neutron
and $^3$H) will have $\Delta T = -1$, and the following emission of a
particle will conserve isospin (and only reach final nuclei with
higher $T$ in the rare \cite{Jon01} $\beta^-$p process). In light
nuclei $\beta^+$-delayed particle emission will mainly proceed in a
similar way --- the key difference is that Fermi decays to the isobaric
analogue state (IAS) are now also energetically possible --- and we
shall mainly consider such systems.

Selection rules for beta transitions in strongly deformed nuclei are
given in \cite{Ala57,boh69}; out of the asymptotic single-particle Nilsson quantum
numbers only $\Omega$ (the angular momentum projection on the
intrinsic symmetry axis) may change by up to one unit for unhindered
transitions, whereas at finite deformation allowed (but hindered,
i.e.\ with some configuration mismatch)
transitions may also take place. These results can of course also be
derived with the methods employed in the next section, and will apply
for the many-body quantum number $K$, the projection of the total
angular momentum on the intrinsic symmetry axis. Results from
charge-exchange reactions indicate \cite{Fuji11} that the $K$-selection
rule is most important for the region of interest here, where
deformation parameters are large, $\delta \approx 0.4$.

We note that deformation in the beta-particle daughter nucleus for
rotational states gives lower-than-usual excitation energies. This may
increase the probability of beta-delayed particle emission to excited
states, allowing richer structure information to be extracted.

The structure of the paper is as follows. In the next section the
general result for particle emission between two deformed nuclei is
derived. Section 3 treats the case of $K=1/2$ and how results turn out
in the particle-rotor model. Section 4 discusses our results in
general, and section 5 outlines the connections to what can be obtained
in transfer reactions. Section 6 applies the formalism to
the decay of $^{21}$Mg and extracts new structure
information. Section 7 gives our summary and conclusions, and the
appendices outline technical relations and definitions.

\section{Problem formulation}

\begin{figure}
\includegraphics[width=0.47\textwidth]{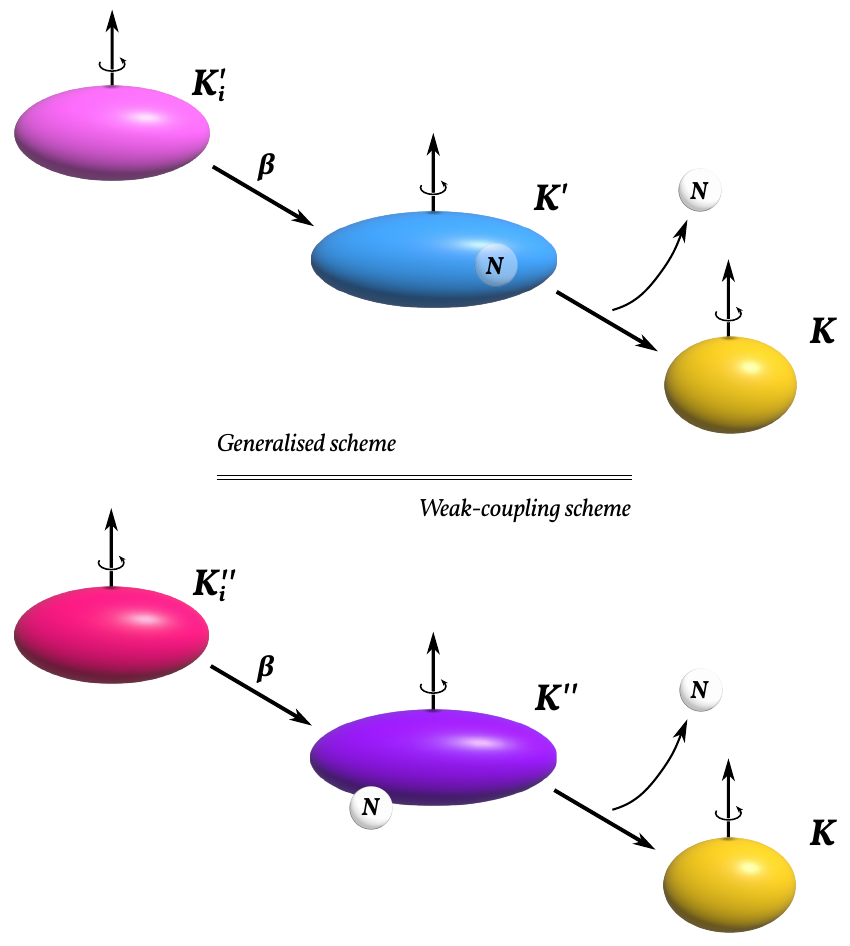}
\caption{ Schematic illustration of the beta-delayed particle emission
  involving rotational states.
  Top: In the general formulation the particle $N$ is emitted from a
  rotating state with many-body quantum number $K'$ that is populated
  in beta-decay. The final state has quantum number $K$.
  Bottom: In the weak-coupling scheme the state populated in
  beta-decay is a rotating state with quantum number
  $K''$ coupled in the laboratory system to the particle. The final
  state is the same.
  }
\label{fig:decayseq} 
\end{figure}

We assume one rotational state in the initial (beta-populated) parent
nucleus, $^{A_i}$XP, with quantum numbers $J'$, $M'$, and $K'$, where
these quantum numbers describe, respectively, total angular
momentum, its projection on the laboratory $z$-axis and on the intrinsic
(symmetry) $z$-axis.  This state emits a particle and populates a
rotational state with quantum numbers $J$, $M$ and $K$ in the daughter
nucleus, $^{A_f}$XD.
The decay sequence is illustrated in the top part of Fig.\ \ref{fig:decayseq}.

\subsection{Initial and final states}

The wave function in the intrinsic body-fixed
system is denoted $\Psi_{K'}(\{\bm{r_i}\})$, where $\bm{r_i}$ stands for
spin, isospin and space coordinate of nucleon $i\in[1,A_i]$
in this coordinate system.
This initial rotational state is then \cite{sie87}
\begin{eqnarray} \label{eq20}
  \int d^3\vec{\omega}' {\cal D}^{J'}_{M'K'}(\vec{\omega}')
 |\Psi_{K'}(\{{\cal R}_{\vec{\omega}'}\bm{r_i}\})\rangle  \; ,
\end{eqnarray}
where ${\cal R}_{\vec{\omega}'}$ is the operator rotating the space and
spin vector, $\bm{r_i}$, successively by the three Euler angles,
$\vec{\omega}'$. Its matrix elements constitute the $D$-function, ${\cal
  D}^{J'}_{M'K'}(\vec{\omega}')$.  The precise definitions are given in
Appendix \ref{appA} together with a number of useful properties.

The structure in Eq.(\ref{eq20}) expresses that the rotational state
is described by a deformed wave function in the intrinsic body-fixed
coordinate system.  Integration over all angles weighted with the
proper $D$-function subsequently restores the correct angular momentum
and its projection on the laboratory coordinate system.

The final state consists of the emitted particle and another nuclear
rotational state.  We describe the emitted particle with coordinate,
$\bm{r}_p$, as a plane wave with wave number, $\bm{k}_p$, multiplied
by the spin function, $\chi_{\Sigma_p}$.  The
particle wave function is then
\begin{eqnarray} \label{eq30}
 \varphi_{\bm{k}_p}(\bm{r}_p) &=& \chi_{\Sigma_p} \exp{(i\bm{k}_p\cdot\bm{r}_p)} \\
  &=&   \chi_{\Sigma_p} 4 \pi \sum_{l,m_l} i^l j_{l}(k_pr_p) Y_{l,m_l}({\hat{\bm{r}}_p})
     Y^{*}_{l,m_l}({\hat{\bm{k}}_p}) \nonumber  \; ,
\end{eqnarray}
where the $\hat{}$ denotes angles only. If the particle is charged, the
plane wave can be replaced by an outgoing Coulomb wave function, or by a
distorted wave function, if an optical potential is used.

The final state rotational wave function formally has the same
structure as Eq.(\ref{eq20}), that is
\begin{eqnarray} \label{eq40}
  \int d^3\vec{\omega} {\cal D}^{J}_{MK}(\vec{\omega})
 |\Psi_{K}(\{{\cal R}_{\vec{\omega}}\bm{r_i}\})\rangle  \; ,
\end{eqnarray}
where the quantum numbers have the same meaning as in Eq.(\ref{eq20}),
and the intrinsic wave function also has the same appearance, but
referring to a smaller set of nucleons. 

\subsection{Overlap matrix element}
Formally, we describe the beta-delayed particle
emission process through the matrix element between the beta-decaying
state and the above final state with the beta decay operator in
between. By inserting a complete set of states in the nucleus $^{A_i}$XP we
end up with a sum over beta matrix elements multiplied with different
overlap matrix elements $ME$ and shall
focus on the latter in the following.

Using Eqs.(\ref{eq20}), (\ref{eq30}) and (\ref{eq40}), the overlap
matrix element, $ME$, between final and initial states is
\begin{eqnarray} \nonumber
 && ME = 4 \pi \sum_{l,m_l} (-i)^l Y_{l,m_l}({\hat{\bm{k}}_p})
  \int \int d^3 \vec{\omega}' d^3\vec{\omega} \\  \label{eq50} 
  && {\cal D}^{J*}_{MK}(\vec{\omega}) {\cal D}^{J'}_{M'K'}(\vec{\omega}')
  \\  \nonumber && \langle \Psi_{K}(\{{\cal R}_{\vec{\omega}}\bm{r_i}\}) |
 \chi^{\dagger}_{\Sigma_p}  j^{*}_{l}(k_pr_p) Y^{*}_{l,m_l}({\hat{\bm{r}}_p})
   | \Psi_{K'}(\{{\cal R}_{\vec{\omega}'}\bm{r_i}\}) \rangle  \; .
\end{eqnarray}

To compute the overlap matrix element, we have to transform all
coordinates to refer to the same frame.  The particle coordinates of both
spin and space, given in the laboratory system, are now rotated to the
intrinsic system of the initial rotating nucleus.  This is done by use
of Eq.(\ref{eqA110}) for each of the tensors,
$\chi^{\dagger}_{\Sigma_p}$ and $Y^{*}_{l,m_l}({\hat{\bm{r}}_p})$, where
the same rotation angle then applies for both cases. Their repective
orders are $s$ and $l$, where $s=1/2$ for nucleons and $0$ for
$\alpha$-particles.  The result is the same set of coordinates, named
$\{\bm{r}'_i\}$, as seen in Eq.(\ref{eqA100}).

The final state coordinates must also be transformed to the same
intrinsic frame of the initial nucleus.  This is done by use of the
two successive rotations in Eq.(\ref{eqA80}), also expressed in terms
of the set $\{\bm{r}'_i\}$, but now rotated by ${\cal
  R}_{\vec{\omega}''}$.  This leaves the final state $D$-function, with
arguments $\omega$, to be expressed in terms of $\omega''$ and
$\omega'$, which is done in Eq.(\ref{eqA90}). Altogether we can
write the overlap in Eq.(\ref{eq50}) as 
\begin{eqnarray} \label{eq60} 
 ME = 4 \pi \sum_{l,m_l} (-i)^l Y_{l,m_l}({\hat{\bm{k}}_p}) \sum_{L,m_l',\Sigma_p'}
 \int \int d^3 \vec{\omega}'' d^3\vec{\omega}' \\ \nonumber  
 {\cal D}^{s*}_{\Sigma_p \Sigma_p'}(\vec{\omega}')
 {\cal D}^{l*}_{m_lm_l'}(\vec{\omega}')
 {\cal D}^{J*}_{ML}(\vec{\omega}') {\cal D}^{J*}_{LK}(\vec{\omega}'')
 {\cal D}^{J'}_{M'K'}(\vec{\omega}') \\ \nonumber
 \langle \Psi_{K}(\{{\cal R}_{\vec{\omega}''}\bm{r_i}'\}) | 
  \chi^{\dagger}_{\Sigma_p'}  j^{*}_{l}(k_pr_p') Y^{*}_{l,m_l'}({\hat{\bm{r}}_p'})
  | \Psi_{K'}(\{\bm{r_i}'\}) \rangle  \; .
\end{eqnarray}
Four of the $D$-functions in Eq.(\ref{eq60}) have the same argument,
$\vec{\omega}'$, which does not appear anywhere else, and consequently
can be integrated out.  Two of them are first combined into one by use of
Eq.(\ref{eqA70}), which naturally introduces the total angular
momentum $j$ of the emitted particle, and the integral over the remaining three
$D$-functions are performed via Eq.(\ref{eqA120}).  The overlap in
Eq.(\ref{eq60}) then becomes
\begin{eqnarray} \nonumber
  && ME=  \sum_{l,m_l,L,m_l',\Sigma_p',j} (-i)^l Y_{l,m_l}({\hat{\bm{k}}_p})
  \langle JMjm_l+\Sigma_p|J'M' \rangle \nonumber \\ \nonumber&&
  \langle JLjm_l'+\Sigma_p'|J'K' \rangle  \langle lm_ls\Sigma_p |jm_l+\Sigma_p \rangle
   \\ \label{eq70} &&
  \langle lm_l's\Sigma_p'|jm_l'+\Sigma_p' \rangle 
  \frac{32 \pi^3}{2J'+1} \int d^3 \vec{\omega}  {\cal D}^{J*}_{LK}(\vec{\omega})
  \\ \nonumber && \langle \Psi_{K}(\{{\cal R}_{\vec{\omega}}\bm{r_i}\}) | 
  \chi^{\dagger}_{\Sigma_p'}  j^{*}_{l}(k_pr_p) Y^{*}_{l,m_l'}({\hat{\bm{r}}_p})
  | \Psi_{K'}(\{\bm{r_i}\}) \rangle  \; .
\end{eqnarray}
One $D$-function is left with the argument $\vec{\omega}''$, now
renamed to $\vec{\omega}$.
If the parity of the initial and final nuclear states is the same (opposite),
only even (odd) values of $l$ are permitted.
The Clebsch-Gordan coefficients imply that the summations are
restricted from initial and final state quantum numbers, that is $m_l
= M'-M-\Sigma_p$, $m_l' = K'-K-\Sigma_p'$, where $\Sigma_p' = \pm 1/2$
and furthermore $j = l \pm 1/2$ for nucleon emission, and
$\Sigma_p'=0$, $j=l$ for alpha particle emission.

\subsection{Normalization}

Before we attempt further reduction of the overlap in Eq.(\ref{eq70}),
we find the normalization factors $N_i$ and $N_f$ for the initial and
final rotational wave functions, respectively. In other words, we must
calculate
\begin{eqnarray} \label{eq80} 
  N_f = \int d^3 \vec{\omega}  {\cal D}^{J*}_{MK}(\vec{\omega})
  \int d^3 \vec{\omega}'  {\cal D}^{J}_{MK}(\vec{\omega}') \\ \nonumber  
  \langle \Psi_{K}(\{{\cal R}_{\vec{\omega}}\bm{r_i}\}) |
  \Psi_{K}(\{{\cal R}_{\vec{\omega}'}\bm{r_i}\}) \rangle \; ,
\end{eqnarray}
which can be reduced by the same technique as used for the overlap in
Eq.(\ref{eq70}). We then easily get 
\begin{eqnarray} \label{eq90} 
 N_f = \frac{8\pi^2}{2J+1} \int d^3\vec{\omega} {\cal D}^{J*}_{KK}(\vec{\omega})  
  \langle \Psi_{K}(\{{\cal R}_{\vec{\omega}}\bm{r_i}\}) |
  \Psi_{K}(\{\bm{r_i}\}) \rangle \; ,
\end{eqnarray}
and $N_i$ follows in complete analogy, related to the other rotational wave
function, where the nucleons belong to the initial state and the
quantum numbers are instead $J'$, $M'$ and $K'$.

Let us assume that the intrinsic states are strongly deformed.
Then the overlaps in the central
pieces of the matrix elements in Eqs.(\ref{eq70}) and (\ref{eq80})
must in the general case be very small unless the rotation angle,
$\theta$, is close to zero \cite{sie87}, that is 
\begin{eqnarray} \label{eq100} 
 && \langle \Psi_{K}(\{{\cal R}_{\vec{\omega}}\bm{r_i}\}) | OP
  | \Psi_{K'}(\{\bm{r_i}\}) \rangle \\ && \nonumber \simeq
 \langle OP \rangle \exp{(iK(\psi+\phi) -\theta^2/\theta_0^2)} \; .
\end{eqnarray}
(This is discussed further in appendix \ref{appB}.)
The average of the operator, $\langle OP \rangle$, is either $1$ from
Eq.(\ref{eq80}) or the function of the emitted particle coordinates in
Eq.(\ref{eq70}).  Using the value of the $D$-function for $\theta=0$
as
 \begin{eqnarray} \label{eq110}
 {\cal D}^{J*}_{LK}(\phi,\theta=0,\psi) = \delta_{LK} \exp{(-iK(\psi+\phi)}) \; ,
\end{eqnarray}
we have for the normalizations
\begin{eqnarray} \label{eq120} 
  N_f \simeq \frac{8\pi^2 (2\pi)^2 \theta_0^2}{2(2J+1)} \;\;, \;\;
  N_i \simeq \frac{8\pi^2 (2\pi)^2 \theta_0^2}{2(2J'+1)} \;.
\end{eqnarray}

\subsection{Normalized overlap}

In analogy to the above we find for the integral in Eq.(\ref{eq70})
\begin{eqnarray} \label{eq130} 
 && \int d^3 \vec{\omega}  {\cal D}^{J*}_{LK}(\vec{\omega})
 \\ \nonumber && \langle \Psi_{K}(\{{\cal R}_{\vec{\omega}}\bm{r_i}\}) | 
  \chi^{\dagger}_{\Sigma_p'}  j^{*}_{l}(k_pr_p) Y^{*}_{l,m_l'}({\hat{\bm{r}}_p})
  | \Psi_{K'}(\{\bm{r_i}\}) \rangle \simeq \\ \nonumber
  &&   \delta_{LK} \frac{1}{2}  (2\pi)^2 \theta_0^2
  \int d^3 \bm{r_p} \chi^{\dagger}_{\Sigma_p'} j^{*}_{l}(k_pr_p)
  Y^{*}_{l,m_l'}({\hat{\bm{r}}_p}) \varphi_{K'}(\bm{r}_p)  \;,
\end{eqnarray}
where we assume that most
of the nucleons in both initial and final states are in the same
rotating state and the initial wave function for the emitted particle
in the intrinsic frame of the initial system is $\varphi_{K'}(\bm{r}_p)$.
(It is discussed in more detail in Appendix \ref{appB}.)

The overlap matrix element from Eq.(\ref{eq70}) with the
normalisations in Eq.(\ref{eq120}) and the reduction in
Eq.(\ref{eq130}), finally gives 
\begin{eqnarray}  \nonumber
  && \frac{ME}{\sqrt{N_i N_f}} = 4 \pi \sqrt{\frac{2J+1}{2J'+1}}
  \sum_{j,l,m_l,m_l',\Sigma_p'} (-i)^l Y_{l,m_l}({\hat{\bm{k}}_p}) \\ \nonumber
  && \langle JMjm_l+\Sigma_p|J'M' \rangle
  \langle JKjm_l'+\Sigma_p'|J'K' \rangle \\ \nonumber
 &&  \langle lm_ls\Sigma_p |jm_l+\Sigma_p \rangle
 \langle lm_l's\Sigma_p'|jm_l'+\Sigma_p' \rangle \\ \label{eq140} 
 && \int d^3 \bm{r_p} \chi^{\dagger}_{\Sigma_p'} j^{*}_{l}(k_pr_p)
  Y^{*}_{l,m_l'}({\hat{\bm{r}}_p}) \varphi_{K'}(\bm{r}_p)  \; .
\end{eqnarray}

Out of the four Clebsch-Gordan coefficients, the first and third
ensure angular momentum conservation in the laboratory system, the
second and fourth the conservation in the intrinsic system. The
selection rule involving the $K$ quantum number is specific to the
deformed nuclei.

The emission rate of the particle is now proportional to the amplitude
in Eq.(\ref{eq140}) squared, summed over final states and averaged over
initial states, i.e.\
\begin{equation}
  \frac{1}{2J'+1} \sum_{M,M'} \frac{|ME|^2}{N_i N_f}
\end{equation}
Furthermore, one must sum or integrate over unobserved quantities,
e.g.\ $\Sigma_p$ and $\hat{\bm{k}}_p$.

\section{The case of a $K=1/2$ system}

The spectra of $K=1/2$ bands in rotating nuclei do not show the usual
$J(J+1)$ energy sequence. Instead the energies fluctuate up and down
more like a $K=1/2$ nucleon (we do not treat alpha particle emission
in this section) that is coupled to a rotating zero angular momentum
core \cite{boh69,sie87}. To test whether particle emission can be
expressed in this framework, we derive the transition matrix element
with the explicit rotator-coupling assumption, and afterwards compare
to the general expression for the transition from a $K=1/2$ initial
structure.

\subsection{Particle-rotor coupling}

In the particle-rotor framework the final state remains the same,
i.e.\ the product of Eqs.(\ref{eq30}) and (\ref{eq40}), whereas the
initial state now is the coupling between a $j=1/2$ nucleon with
wave function in the laboratory coordinate system (i.e.\ the weak
coupling limit \cite{sie87}; we separate out the spin, angular and radial
wave functions), and the rotating core with quantum numbers
$J''M''K''$, see also the bottom part of Fig.\ \ref{fig:decayseq}:
\begin{eqnarray} \nonumber  
 && \sum_{M'',m_j'',m_l'',\Sigma''}\langle J''M'' \frac{1}{2} m_j'' | J'M'\rangle
  \langle l'' m_l'' \frac{1}{2} \Sigma''| \frac{1}{2} m_j'' \rangle
 \\ \nonumber
 &&  Y_{l'',m_l''}({\hat{\bm{r}}_p}) \chi_{\Sigma''} \varphi_{1/2}(r_p)
 \\  \label{eq160}   &&
 \int d^3\vec{\omega}' {\cal D}^{J''}_{M''K''}(\vec{\omega}')
  |\Psi_{K''}(\{{\cal R}_{\vec{\omega}'}\bm{r_i}\})\rangle \; .
\end{eqnarray}

To compute the overlap matrix element, we use Eqs.(\ref{eq30}), (\ref{eq40})
and (\ref{eq160}). 
\begin{eqnarray} \nonumber
  && ME = 4 \pi \sum_{l,m_l,M'',m_j''} (-i)^l Y_{l,m_l}({\hat{\bm{k}}_p})
 \langle J''M'' \frac{1}{2} m_j'' | J'M'\rangle
   \\  \label{eq170} &&  
  \langle l m_l \frac{1}{2} \Sigma_p| \frac{1}{2} m_j'' \rangle
  \int \int d^3 \vec{\omega}' d^3\vec{\omega} 
   {\cal D}^{J*}_{MK}(\vec{\omega}) {\cal D}^{J''}_{M''K''}(\vec{\omega}')
   \\  \nonumber && \langle \Psi_{K}(\{{\cal R}_{\vec{\omega}}\bm{r_i}\}) |
   \varphi_{1/2}(r_p)  j^{*}_{l}(k_pr_p)
  |\Psi_{K''}(\{{\cal R}_{\vec{\omega}'}\bm{r_i}\})\rangle \; ,
\end{eqnarray}
where we combined the spin and orbital wave functions,
$\chi_{\Sigma''}$ and $\chi_{\Sigma_p}$, and the orbital,
$Y_{l'',m_l''}$ and $Y_{l,m_l}$ in initial and final states to give
the $\delta$-functions, $\delta_{\Sigma_p,\Sigma''}$, $\delta_{l,l''}$
and $\delta_{m_l,m_l''}$.  We continue to reformulate the integrations
in Eq.(\ref{eq170}) as was done above in going from Eq.(\ref{eq50}) to
Eq.(\ref{eq60}).  The nucleon's spin and orbital wave functions
disappear, and we get the similar, but simpler expression
with only three $D$-functions
\begin{eqnarray} \nonumber
 ME = \sum_{l,m_l,L,M'',m_j''} 4\pi (-i)^l Y_{l,m_l}({\hat{\bm{k}}_p})
   \\ \nonumber  
 \langle J''M'' \frac{1}{2}m_j''|J'M'\rangle 
 \langle l m_l \frac{1}{2} \Sigma_p| \frac{1}{2} m_j'' \rangle
 \\ \nonumber    \int \int d^3 \vec{\omega}'' d^3\vec{\omega}'
 {\cal D}^{J*}_{ML}(\vec{\omega}') {\cal D}^{J*}_{LK}(\vec{\omega}'')
 {\cal D}^{J''}_{M''K''}(\vec{\omega}') \\  \label{eq180} 
 \langle \Psi_{K}(\{{\cal R}_{\vec{\omega}''}\bm{r_i}'\}) | j^{*}_{l}(k_pr_p) 
 \varphi_{1/2}(r_p)  | \Psi_{K''}(\{\bm{r_i}'\}) \rangle.
\end{eqnarray}
The two $D$-functions with the same argument, $\vec{\omega}'$, are
integrated to give the $\delta$-functions, $\delta_{J,J''}$,
$\delta_{M,M''}$ and $\delta_{L,K''}$, that is

\begin{eqnarray} \nonumber
 && ME=  \sum_{l,m_l,m_j''} 4 \pi (-i)^l Y_{l,m_l}({\hat{\bm{k}}_p}) 
 \\ \nonumber && \langle J M \frac{1}{2} m_j''|J'M' \rangle 
 \langle l m_l \frac{1}{2} \Sigma_p| \frac{1}{2} m_j'' \rangle
  \\ \nonumber &&  \frac{8 \pi^2}{2J+1} 
  \int d^3 \vec{\omega}  {\cal D}^{J*}_{K''K}(\vec{\omega})
  \langle \Psi_{K}(\{{\cal R}_{\vec{\omega}}\bm{r_i}\}) | 
  \Psi_{K''}(\{\bm{r_i}\}) \rangle \\ \label{eq190}  &&
 \int d^3\bm{r_p}   j^{*}_{l}(k_pr_p)  \varphi_{1/2}(r_p) \; .
\end{eqnarray}
Again, one $D$-function remains with the argument $\vec{\omega}''$, now
renamed to $\vec{\omega}$.  The assumptions of strong deformation and
of identical initial and final state
rotating cores, finally gives
\begin{eqnarray} \nonumber
  && \frac{ME}{\sqrt{N_iN_f}} =  4 \pi
  \sum_{l,m_l,m_j''}  (-i)^l Y_{l,m_l}({\hat{\bm{k}}_p})
  \langle J M \frac{1}{2} m_j''|J'M' \rangle  \\ &&   \label{eq200}
  \langle l m_l \frac{1}{2} \Sigma_p| \frac{1}{2} m_j'' \rangle
  \int d^3\bm{r_p}  j^{*}_{l}(k_pr_p) 
  \varphi_{1/2}(r_p) \; .
\end{eqnarray}

This derivation explicitly highlights the assumption that the
rotating cores are identical before and after the nucleon emission.
Depending on whether there is a change or not in parity between the
initial and final states, the $l$-values will be odd or even;
this implies that $l=0$ or $l=1$ when the nucleon has total
angular momentum of $j=1/2$.
Then Eq.(\ref{eq200}) simplifies for $l=0$ to 
 \begin{eqnarray}   \label{eq200a}
  && \frac{ME}{\sqrt{N_iN_f}} = \\ \nonumber &&
   \sqrt{4 \pi}
  \langle J M \frac{1}{2} \Sigma_p |J'M' \rangle
   \int d^3\bm{r_p}  j^{*}_{0}(k_pr_p)  \varphi_{1/2}(r_p) \; .
\end{eqnarray}
For $l=1$ we are again left with only one term, that is
\begin{eqnarray} \nonumber
  && \frac{ME}{\sqrt{N_iN_f}} =  - 4 \pi i
  Y_{1,m_l}({\hat{\bm{k}}_p})
 \langle J M \frac{1}{2} m_j''|J'M' \rangle  \\ &&   \label{eq200b}
 \langle 1 m_l \frac{1}{2} \Sigma_p| \frac{1}{2} m_j'' \rangle
 \int d^3\bm{r_p}  j^{*}_{1}(k_pr_p) \varphi_{1/2}(r_p) \; ,
\end{eqnarray}
since $m_j'' = M' - M$ and $m_l = M' - M - \Sigma_p $.  If desired,
the Clebsch-Gordan coefficients can be explicitly inserted, where we
get
\begin{eqnarray} \label{eq200c}
  \langle 1 m_l \frac{1}{2} \Sigma_p| \frac{1}{2} m_j'' \rangle 
 =  (-1)^{1/2+\Sigma_p} \sqrt{\frac{1-2 \Sigma_p m_l}{3}} \;
\end{eqnarray}
with $m_l = M' - M - \Sigma_p $.

\subsection{Special particle-rotor model results}

Eq.(\ref{eq200a}) becomes simpler for transitions to the ground state,
where $J=K=M=0$ and the parity is even, in which case
we get $J'=1/2$, $M'=\Sigma_p=\pm 1/2$, and the expression:
\begin{eqnarray}   \label{eq200d}
  && \frac{ME}{\sqrt{N_iN_f}} =  \sqrt{4 \pi}
   \int d^3\bm{r_p}  j^{*}_{0}(k_pr_p)  \varphi_{1/2}(r_p) \; .
\end{eqnarray}
This means that only $\frac{1}{2}^+$ states can decay to the ground
state via $s$-wave emission.

Whereas in the general case the emitted particle is described in the
intrinsic system of the initial nucleus, it is in the weak coupling
limit of the particle-rotor model described in the laboratory
system.
It is then of interest to see whether the final results are
consitent. 

The general formula in Eq.(\ref{eq140}) will in the case of $K'=1/2$
and $l=0$ simplify: As $s=1/2$ we have $j=1/2$, and two of the
Clebsch-Gordan coefficients are unity, so 
\begin{eqnarray}  \nonumber
  && \frac{ME}{\sqrt{N_i N_f}} = \sqrt{4 \pi} \sqrt{\frac{2J+1}{2J'+1}}
  \sum_{\Sigma_p'}  \\ \nonumber
  && \langle JM\frac{1}{2}\Sigma_p|J'M' \rangle
  \langle JK\frac{1}{2}\Sigma_p'|J'K' \rangle \\ \label{eq140a} 
 && \int d^3 \bm{r_p} \chi^{\dagger}_{\Sigma_p'} j^{*}_{0}(k_pr_p)
  Y^{*}_{0,0}({\hat{\bm{r}}_p}) \varphi_{K'}(\bm{r}_p)  \; .
\end{eqnarray}
The wavefunction $\varphi_{K'}(\bm{r}_p)$ has a radial part, an
angular part, which must have $l=0$ to make the integral nonvanishing,
and a spin part. A closer inspection shows that
$\varphi_{K'}(\bm{r}_p)$ wil have two terms (with quantum numbers $\pm
K'$). This more general structure is discussed in Appendix \ref{appB}
where the equivalence of the two formulations is shown explicitly for
$K=0$.

\section{General results}

This section discusses specific results of the general formalism
for emission of nucleons and $\alpha$ particles as well as the
selection rules in $J$ and $K$.

\subsection{Selection rules}
The Clebsch-Gordan coefficients in Eq.(\ref{eq140}) imply that
\begin{equation}  \label{eq205}
  \vec{J}+\vec{j}=\vec{J'} \;,\; |M-M'| \leq j \;,\; |K-K'| \leq j \;,
\end{equation}
corresponding to the conservation of angular momentum and that the
difference in $M$ and $K$ values in the initial and final nucleus
cannot differ by more than $j = l \pm s$. The interesting selection
rule comes from the last term involving $K$, since all members of a
rotational band will have the same value of $K$, but increasing values
of $J$ (and $M$). A difference in $K$-values can therefore enforce a
higher value of $j$ (and thereby of $l$) than allowed by angular
momentum conservation alone.
As a concrete example, only a $K'=1/2$ band can emit s-wave
nucleons to a $K=0$ band.

For states that are band heads the $J$ and $K$ quantum numbers are the
same (except possibly for $K$ or $K'=1/2$ bands) in which case there
are no extra restrictions from the $K$-selection rule, but emission from
(or to) levels higher in a band may be affected by it.

\subsection{Emission of protons or neutrons}
Nucleons have $s=1/2$, so that given a value of $l$ there are two
possible $j$-values. However, due to parity conservation, given a
value of $j$ only one of the two orbital angular momentum values, $l=j \pm
1/2$, is permitted.

The probability of particle emission is given in Eq.(\ref{eq140}) and
depends (as usual) partly on the overlap matrix element, partly on
the penetrability factor that is related to the square of the Bessel
function (or Coulomb wave function for charged emitted particles),
cf.\ chapters 7.4.1 and 10.2 in \cite{Tho09}.

The first special case of interest is emission of nucleons to the
ground state of an even-even nucleus. Since $J=M=K=0$ this reduces
Eq.(\ref{eq140}) significantly, as $j = J'$, $m_l+\Sigma_p=M'$ and
$m_l'+\Sigma_p'=K'$. The total expression reduces to
\begin{eqnarray}  \nonumber
  && \frac{ME}{\sqrt{N_i N_f}} = 4 \pi \sqrt{\frac{1}{2J'+1}}
  \sum_{l,\Sigma_p'} (-i)^l Y_{l,m_l}({\hat{\bm{k}}_p}) \\ \nonumber
  && 
  \langle lm_l \frac{1}{2}\Sigma_p |J'M' \rangle
   \langle lm_l' \frac{1}{2}\Sigma_p'|J'K' \rangle \\ \label{eq210} 
 && \int d^3 \bm{r_p} \chi^{\dagger}_{\Sigma_p'} j^{*}_{l}(k_pr_p)
  Y^{*}_{l,m_l'}({\hat{\bm{r}}_p}) \varphi_{K'}(\bm{r}_p)  \; ,
\end{eqnarray}
The quantum numbers, $M'$ and $K'$, have given values characterizing the initial
state. The spin-projection, $ \Sigma_p = \pm 1/2$, is in principle an
observable in the final state characterizing the nucleon, which implies
that $m_l=M'-\Sigma_p$.  In contrast,
the spin projection, $\Sigma_p'$, of the nucleon in the initial state,
$\Sigma_p' = \pm 1/2$, is still a summation index, which however
implies $m_l' = K' - \Sigma_p'$.

As discussed in detail in the next section
this expression is very similar to the one of transfer reactions on an
even-even nucleus \cite{Elb69}. Emission to higher states in the ground state band
gives a more complex expression.

The second special case is that of emission of an $l=0$ nucleon
independent of what final state is reached. Clearly we have
$m_l=m_l'=0$ and $j=s=1/2$.
The simpler expression becomes
\begin{eqnarray}  \nonumber
  && \frac{ME}{\sqrt{N_i N_f}} =  \sqrt{\frac{2J+1}{2J'+1}} \sum_{\Sigma_p'}
  \langle JM\frac{1}{2}\Sigma_p|J'M' \rangle \\ \label{eq230}  &&
  \langle JK \frac{1}{2}\Sigma_p'|J'K' \rangle
   \int d^3 \bm{r_p} \chi^{\dagger}_{\Sigma_p'} j^{*}_{0}(k_pr_p)
   \varphi_{K'}(\bm{r}_p)  \; .
\end{eqnarray}
For emission to a $K=0$ band the sum reduces to the term with
$\Sigma_p' =K'$ and we recover the previous result that only a
$K'=1/2$ band can emit angular momentum zero protons to a $K=0$ 
band.

\subsection{Emission of $\alpha$-particles}

For $\alpha$-particles $s=0$ and $j=l$. Inserting this in
Eq.(\ref{eq140}) gives a significant simplification, i.e.\

\begin{eqnarray}  \nonumber
  && \frac{ME}{\sqrt{N_i N_f}} = 4 \pi \sqrt{\frac{2J+1}{2J'+1}}
  \sum_{l,m_l,m_l'} (-i)^l Y_{l,m_l}({\hat{\bm{k}}_p}) \\ \nonumber
  && \langle JMlm_l|J'M' \rangle
  \langle JKlm_l'|J'K' \rangle \\  \label{eq280} 
 && \int d^3 \bm{r_p}  j^{*}_{l}(k_pr_p)
  Y^{*}_{l,m_l'}({\hat{\bm{r}}_p}) \varphi_{K'}(\bm{r}_p)  \; .
\end{eqnarray}

Emission of s-wave $\alpha$-particles must take place to states with
exactly the same $J$, $M$ and $K$ as the original state and the
normalized matrix element reduces to the following overlap integral
\begin{eqnarray}  
  && \frac{ME}{\sqrt{N_i N_f}} = \label{eq290} 
 \int d^3 \bm{r_p}  j^{*}_0(k_pr_p)
  \varphi_{K'}(\bm{r}_p)  \; .
\end{eqnarray}

\section{Relation to transfer reactions}
\begin{figure}
\includegraphics[width=0.47\textwidth]{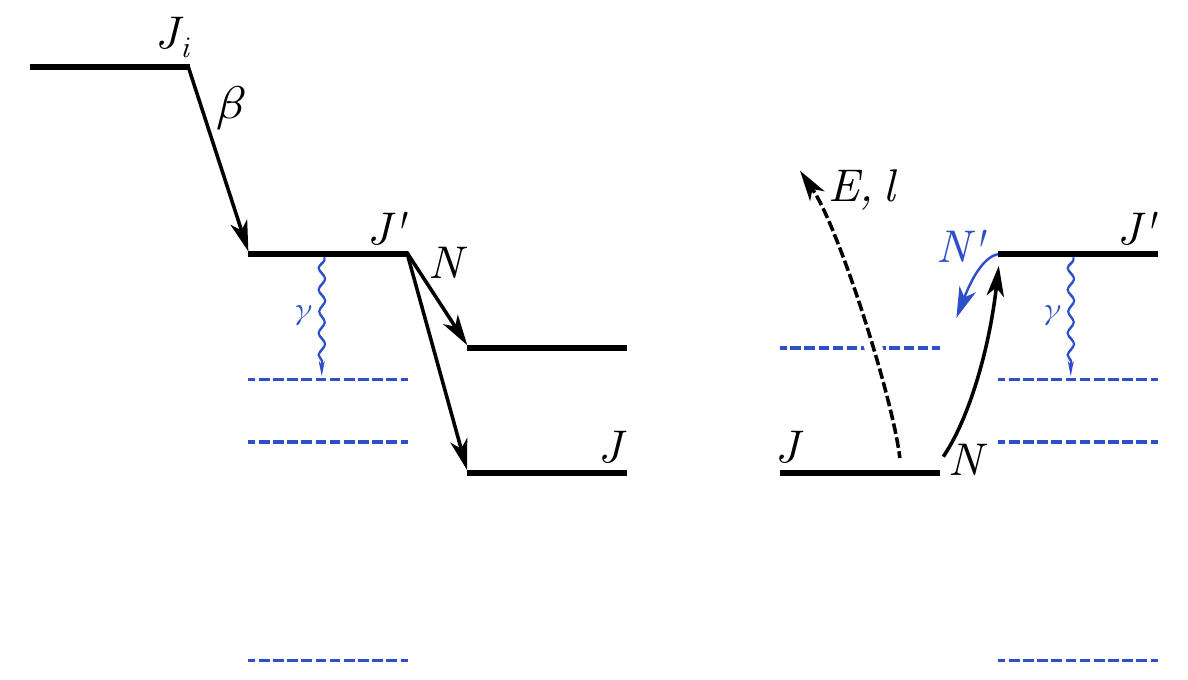}
  \caption{ Left: Beta-delayed particle emission where the particle $N$
    is emitted from a level with spin $J'$ to a level with spin $J$.
    Observables can be the energy and direction of the particle as well as
    the beta particle.
    Right: Transfer reaction from a level with spin $J$ to a level
    with spin $J'$. Typical observables are the energy and angular
    distribution of the outgoing projectile-like fragment.
  }
\label{fig:beta_trans} 
\end{figure}

We have so far mainly discussed the particle emission process that can
follow a beta-decay, see left hand side of Figure \ref{fig:beta_trans}.
As shown above, structure information for the emitting state can be
derived from the decay pattern of the level. If the preceding beta
particle is also detected, one can (provided the angular momenta of
the participating states and the emitted particle is 1$\hbar$ or larger)
derive further constraints from the angular distribution between the
beta particle and the emitted particle. There is also information from
the beta decay selection rules (as mentioned in the introduction).

It may be enlightening to compare this to the information
traditionally extracted from single-particle transfer reactions, see
right hand side of Figure \ref{fig:beta_trans}. A particle is
transferred from a projectile to the target (with spin $J$) populating
a level with spin $J'$, i.e.\ the reverse reaction to the particle
emission. Observation of the energy and angular distribution of the
projectile fragment allow deduction of the excitation energy of the level and the
angular momentum of the transferred particle. Note that levels both
above and below the particle threshold in the final nucleus may be
populated, but that only transitions from the ground state of the
initial nucleus are probed.
(If the experimental set-up allows, the decay of the level can of
course also be recorded, in which case the formalism presented in this
paper applies.)

Transfer reactions are often used to extract spectroscopic factors for
single-particle configurations. The many-body matrix element in our
Eq.(\ref{eq70}) is closely related to the overlap functions that are
used to define the spectroscopic factors, cf.\ chapter 5.3 in
\cite{Tho09}. If we do not make the assumption in section II.D that the
wave function of the level is completely described as a single
particle outside the final state, we see that population of
the level and the decay of it reveals the same physics information (as
is to be expected). 

The selection rule in Eq.(\ref{eq140}) involving the $K$ quantum
number is specific to the deformed nuclei; a similar term occurs for
transfer reactions that involve deformed nuclei, see e.g.\ section 4
in \cite{Elb69} and references therein.  In the integral in
Eq.(\ref{eq140}) the spin and angular momentum parts of the wave
function of the outgoing particle will select the parts of
$\varphi_{K'}(\bm{r}_p)$ with the corresponding angular momentum; this
is again similar to the overlap occurring for transfer reactions and
gives rise to what is often named \cite{Elb69,boh69} a ``fingerprint''
signal for the structure of the state.  For transfer reactions this
has been used to probe the structural details of Nilsson model states
by populating different excited states in a rotational band.  As the
example in the next section shows, we can extract structure information for each
state that emits particles to several final states and may possibly
extract more detailed information
 through comparisons with theoretical
structure calculations.

\section{The decay of  $^{21}\mathbf{Mg}$}

The practical use of the results will be
illustrated with the decay of $^{21}$Mg, where many aspects of
the use of the formalism can be seen in a somewhat simple setting,
since much of the structure of nuclei in the lower sd-shell can be
discussed in terms of s-wave and d-wave nucleons, only.
A brief overview of other possible cases is given in the last subsection.

\begin{figure}
\includegraphics[width=0.48\textwidth]{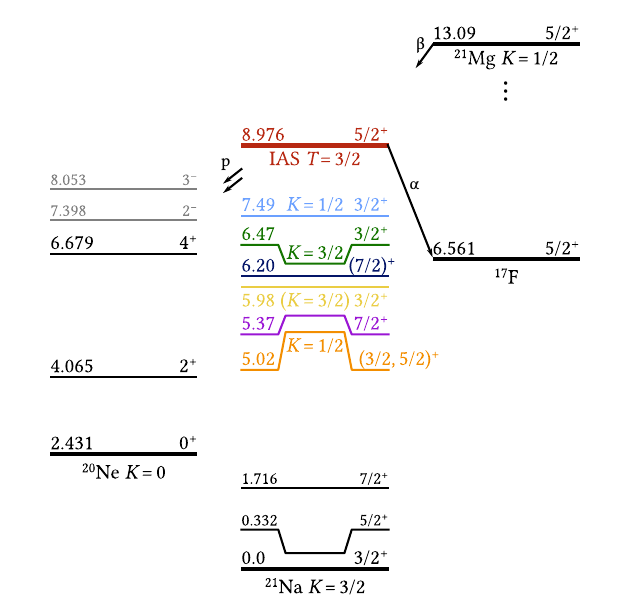}
  \caption{ A sketch of some of the levels that occur in the
    beta-delayed proton decay of $^{21}$Mg \protect\cite{Jen24}. Several
    of the levels in $^{21}$Na that are mentioned in the
    text are displayed with the colour coding being the same as in Figure
    \protect\ref{fig:particle}.  
  }
\label{fig:levels} 
\end{figure}

The ground state of $^{21}$Mg with one neutron in the Nilsson orbital
[220]1/2$^+$ (the single-particle quantum numbers are [$Nn_z \Lambda$]$\Omega^P$) 
has spin-parity 5/2$^+$ and is expected to be member of a
$K=1/2$ band with large positive decoupling parameter, as its
mirror nucleus $^{21}$F \cite{boh69,Ell68} (see also \cite{Kim11,Sun24} for
independent structure calculations). The isospin is 3/2 and the same quantum numbers
apply for the Isobaric Analogue State (IAS) at 8.97 MeV excitation
energy in $^{21}$Na, see Figure \ref{fig:levels}.

Allowed beta-decay of $^{21}$Mg populates several excited states in
$^{21}$Na with $J'=3/2,5/2,7/2$, positive parity and $K' = 1/2,3/2$; those above 3.5
MeV excitation energy decay mainly by emiting protons to states in
$^{20}$Ne, at first only to the ground state but above 5 MeV
excitation energy also to excited states. Most states fed in $^{20}$Ne
belong to the ground state rotational band of $K=0$ and $J=0,2,4$ with
positive parity. A recent paper \cite{Jen24} gives an overview of the
current experimental knowledge.

The general formalism given above will be used to set restrictions on
the quantum numbers of the observed levels and thereby expand the
current knowledge on 
the rotational bands in $^{21}$Na (see \cite{boh69,Bij21} for established
structure at lower excitation energy). Eq.(\ref{eq140}) gives two
general restrictions, the first arises from the Clebsch-Gordan
coefficients (the selection rules) that imply that states with
$K'=1/2$ may emit both s-wave and d-wave protons, whereas
$K'=3/2$ may only emit d-wave protons.

The second restriction comes from consideration of the wave functions
in the integral in Eq.(\ref{eq140}). The internal single-particle wave
function, $\varphi_{K'}(\bm{r}_p)$,
mainly contains s- and d-waves (we are in the sd-shell) and the integral will
therefore restrict the $l$-values to 0 and 2. A $J'=7/2$ level will then not
decay to the ground state, but can emit protons to the $2^+$ (and $4^+$)
levels.
The $J' = 3/2, 5/2$ levels can decay to the ground state by emitting a
d-wave proton and to the $2^+$ level by d- or s-wave, but s-wave is
only allowed for $K'=1/2$.

\subsection{Proton emission results}

\begin{figure*}
\includegraphics[width=\textwidth]{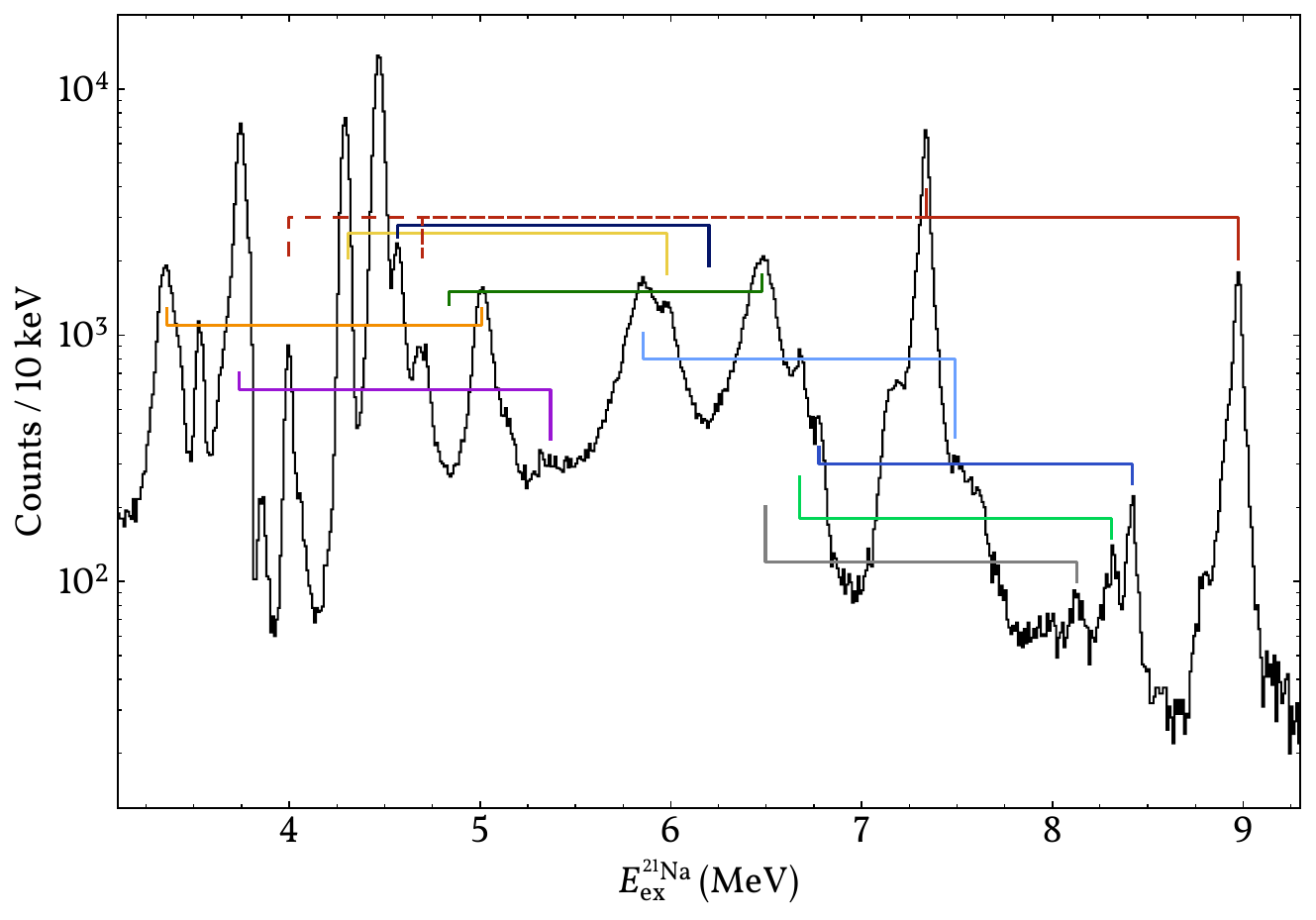}
  \caption{The beta-delayed proton spectrum recorded in
    \protect\cite{Jen24} shown versus the excitation energy $E_{ex}$ in
    $^{21}$Na (deduced assuming the proton is emitted to the $^{20}$Ne
    ground state). The solid lines connect peaks that originate from the same
    level, to the right the transition to the ground state, to the left
    to the excited $2^+$ state in $^{20}$Ne (displaced 1.634 MeV
    downwards).  
    The IAS at 8.97 MeV decays also to higher excited states, this is indicated
    with the dashed line. Below $E_{ex} \approx$ 4.2 MeV the shown
    intensities are displaced by factors 2--3, see \protect\cite{Jen24}
    for details.
  }
\label{fig:particle} 
\end{figure*}

An overview of the current experimental knowledge on the structure of $^{21}$Na can
be found through the papers \cite{Fir15,Jen24}. Parts of the pertinent
information on the levels that appear in the decay are reproduced in
Table \ref{tab:peaks}, and the observed proton spectrum from
\cite{Jen24} is shown in Figure \ref{fig:particle}. The figure shows
the singles proton spectrum and many excited levels could
energetically decay both to the $0^+$ ground state and $2^+$ first
excited state in $^{20}$Ne; for each level the position of the protons
from these two transitions is marked, and Table \ref{tab:peaks}
includes the observed branching ratios $b_0$ and $b_2$ of the two
transitions as well as the penetrability for s- and/or d-wave emission
for the transitions. For ease of reference, the most relevant parts
of the decay scheme are reproduced also in Figure \ref{fig:levels}.

The penetrabilities are relevant for two different aspects: (1) in the wave
function overlap integral
the penetrability gives the approximate value of how much the
continuum wave function is suppressed by the Coulomb and angular
momentum barriers;
(2) we may estimate the width of a level via R-matrix theory (see
e.g.\ chapter 10 in \cite{Tho09}) as twice the
penetrability times the reduced width, an upper limit of which is the
so-called Wigner width, which here is about 1.6 MeV.
The ratio of penetrabilities therefore enters in the ratio $b_2/b_0$ and
levels with large width $\Gamma$ must have high values both
of the penetrability and the wave function amplitude. 
This may be relevant for the levels at 5.02 MeV, 6.47 MeV and 7.49 MeV.

\begin{table*}
\caption{Beta-delayed protons from $^{21}$Na. Results from the
  beta-decay experiment in \protect\cite{Jen24} to the left,
  literature values from \protect\cite{Fir15} in the middle, and
  penetrabilities and deduced $J'$ and $K'$ values to the
  right. $E_{ex}$ is the excitation energy in $^{21}$Na, $\Gamma$ the
  level width and $b_p$ the branching ratio.}
\label{tab:peaks} 
\begin{tabular}{lll|lll|ccccl}
\hline\noalign{\smallskip}
$E_{ex}$ & \multicolumn{2}{c}{b$_{p}$ (\%)$^a$}  & $E_{ex}$(lit) & $\Gamma$(lit)
  & $J'^{\pi}$ & \multicolumn{3}{c}{Penetrability} & $J'$ & $K'$ \\
(MeV) & $0^+$ & $2^+$ & (keV) & (keV) & & gs, $l=2$ & $2^+$, $l=0$ &
                                                                    $2^+$, $l=2$ & & \\
\noalign{\smallskip}\hline\noalign{\smallskip}
 5.02(1) &  4.6(3) & 4.4(4) &  & 110(15)$^b$ & & 0.134 & 0.061 &
                                        0.0016 & 3/2, 5/2 & 1/2 \\
 5.37(1) &  $<0.4$ & 10.8(4) &  & & & 0.201 & 0.186 & 0.008 & 7/2 & \\   
 5.98(2) &  0.15(5) & $<0.05$ & & & & 0.344 & 0.488 & 0.044 & 3/2$^b$ & (3/2) \\
 6.20(1) &   & 1.5(4) & & & & 0.402 & 0.603 & 0.068 & 7/2 & \\
 6.47(2) &  6.5(6) & 0.5(3) & 6468(20) & 145(15) & $3/2^+$ & 0.476 &
                                                                     0.740 & 0.104 & & 3/2 \\
 7.49(2) &  $<1.4$ & 7.2(6) & 7609(15) & 112(20) & $3/2^+$ & 0.777 &
                                                                     1.203 & 0.313 & & 1/2 \\
 8.13(2) &  0.16(3) & 0.5(3) & 8135(15) & 32(9) & $5/2^+$ & 0.970 &
                                                                    1.453 & 0.484 & & \\
 8.31(2) &  0.20(2) & 0.6(2) & 8397(15) & 30(13) & $3/2^+$ & 1.024 &
                                                                     1.518 & 0.535 & &  \\
 8.42(2) &  0.23(2) & 0.4(2) & 8464(15) & 25(9) & $3/2^+$ & 1.057 &
                                                                    1.557 & 0.567 & & \\
 8.97(1) &  2.10(3) & 8.4(3) & 8976(2) & 0.65(5) & $5/2^+$ & 1.219 &
                                             1.742 & 0.730 & & 1/2 \\
\noalign{\smallskip}\hline
\end{tabular}

$^a$The fraction of the total beta-delayed proton spectrum.

$^b$Value deduced in \protect\cite{Jen24}.
\end{table*}

Table \ref{tab:peaks} also gives the deduced quantum numbers appearing from the
following analysis.

The first level in $^{21}$Na to emit protons to the $2^+$ state is at
5.02 MeV (orange in Figs \ref{fig:levels} and \ref{fig:particle}).
It has about the same branching ratio to the two states in
$^{20}$Ne which strongly suggests that it proceeds via s-wave emission
to the $2^+$ state. It must therefore have $K' = 1/2$. Since protons
are emitted to the ground state, the spin cannot be 7/2.

The next level
at 5.37 MeV (purple) has a $b_2$ more than 20 times larger than $b_0$, so the
latter decay is strongly hindered. It must therefore have $J' = 7/2$,
which is consistent with the assignment in \cite{Bij21}.
In a similar way, the 6.20 MeV level (dark blue) is only observed to decay to the excited state, and
must also have $J'=7/2$.

The 5.98 MeV level (yellow) is assigned in \cite{Jen24} as $J'=3/2$ and is only
seen to decay to the ground state. The excited state transition
would lie below one of the strongest proton peaks and is weaker by
more than a factor 3, which could indicate that $K'$ cannot be
1/2; it must then be 3/2. A similar argument applies for
the 6.47 MeV level (dark green) that is known to have $J'=3/2$ and has a $b_0$ more than
an order of magnitude larger than $b_2$. This indicates that both
proton transitions occur with $l=2$ and implies that the level has $K'=3/2$.

The 7.49 MeV level (light blue) is also $J' = 3/2$, but decays mainly (by a factor
more than 5) to the excited $2^+$ state. That points to $K'=1/2$ and a
rather high s-wave content in its wave function. In the same way one
can argue that the $J' = 5/2$ IAS at 8.97 MeV (red) has $K' = 1/2$, as is
expected since it is the analogue of the $^{21}$Mg ground state.

The three levels at 8--8.5 MeV (grey, light green and blue) are all observed to decay to both
states in $^{20}$Ne (consistent with their literature spin
assignments), but the accuracy of their branching ratios is too 
low to give information on their $K$-values.

\subsection{Structural implications}

A suggested assignment of Nilsson single-particle configurations to the lowest bands
in $^{21}$Ne is given in Table 5-8 in \cite{boh69} and should be very
similar to the one of $^{21}$Na. Assignments for $K$-values for both
nuclei are given in \cite{Bij21} and differ mainly in the
interpretation of the lowest $K'=1/2$ band that is attributed to
[211]1/2 in \cite{boh69} and rather to a core excited configuration
(e.g.\ [220]1/2) in \cite{Bij21}, note that these two configurations
could be expected to lie close in excitation energy and therefore
could be susceptible to Coriolis mixing.
Apart from this and the [211]3/2 ground state band, both sources
attribute two negative parity bands to core excitations ([101]1/2 and
[330]1/2). This exhausts the structure up to 4--5 MeV \cite{Fir15,Bij21}.

The remaining single-particle configurations are [202]5/2, [200]1/2
(most likely with a small decoupling parameter) and [202]3/2. It is
important to note that more complex configurations will come from
exciting protons or neutrons from the [220]1/2 orbit (a clear example
is the IAS that furthermore has a different isospin value), but these
can not be described with our simplest model ansatz.  The
simplification in Eq.(\ref{eq130}) no longer holds and we will now
need to evaluate the more complex integral in Eq.(\ref{eq70}), but
the selection rules are still very similar. The proton decay of the levels is likely
to be hindered and the levels would naturally become narrow.

We do not expect sizable beta-decay to the [202]5/2 configuration
due to the K-difference of 2. This is at variance with the
interpretation in \cite{Bij21} of the $5/2^+$ level at 4294 keV and
the $7/2^+$ level at 5.37 MeV as belonging to a $K' = 5/2$ band. We would
rather expect them to be related in a $K' = 3/2$ band.

Using the estimated energies of the configurations from \cite{Bij21}
one could expect the [202]3/2 configuration to correspond to the
$K'=3/2$ band that starts at the 6.47 MeV level, and the [200]1/2
configuration to start slightly lower in energy (but note that it may
mix with the other $K'=1/2$ bands).
The wide 5.02 MeV level is more likely to be a member of the [211]1/2 band,
but the 7.49 MeV level could belong to the [200]1/2 band that has a
relatively large s-wave component cf.\ Table 5-9 in \cite{boh69}.
A large level width is due to either s- or d-wave strength.

The IAS in itself cannot proton decay due to isospin conservation,
but it is likely to mix with two other close-lying levels
\cite{Jen24}. One of these is rather broad and may be related to 
the 7.49 MeV level, this would then facilitate the proton decays to
the ground state band in $^{20}$Ne. The decays to the $2^-$ band
(transitions are known to both the $2^-$ and $3^-$ levels) may then be
due to mixing with the other level.
Comparing the branching ratios and penetrabilities to the different
final states \cite{Jen24} one observes a structural preference for
decays to the $2^-$ and $4^+$ states.
As a potential signature, we note 
that the broad component around the IAS seems present in the $4^+$ gated spectrum,
but not in the $2^-$ gated spectrum in \cite{Jen24}. More statistics
would be needed to resolve this.

We note that the
simple particle-rotor picture for the beta decay lets the $1/2^+$ state
in the intrinsic frame go to either $1/2^+$ or $3/2^+$ (these then
become the $K'$-values) to which the core rotation must be added.
The $3/2^+$ ground state in a $K'=3/2$ band may therefore be hindered in beta
feeding compared to the $5/2^+, 7/2^+$ members, which fits with the
skew beta feeding pattern observed to the ground state band (Table 3 in
\cite{Jen24}). 

The preference for proton emission to the final $2^+$ rotational state
seems related to the fact that the initial state has a core in a $2^+$
rotation. The beta decay will preferentially go to $^{21}$Na states
that overlap with the rather simple structure of the Gamow-Teller
Giant Resonance (GTGR, a spin- and isospin-flip of the $^{21}$Mg
ground state). The proton emission of these states select components
with composition one proton plus a $^{20}$Ne state; as the GTGR lies
at a somewhat high energy, many components of the states fed in the
beta decay can be expected to have a rather complex structure, so it
is not that surprising if the simple rotational component fed in beta
decay overlaps with the simple rotational component preferred in the
proton emission.

\subsection{Emission of $\alpha$-particles}

In the case of $^{21}$Na the final state after emission of an
$\alpha$-particle is $^{17}$F, which is not deformed. The formalism
will therefore not apply, but the main physics ingredients in the
particle emission will be the conservation of angular momentum and an
overlap matrix element, so the difference to the deformed case mainly
lies in the Clebsch-Gordan coefficient with the $K$ quantum number.  The
observed $\alpha$-decays stem from levels around the IAS, which as
discussed above have a more complex structure.

\subsection{Particle emission from other sd-shell nuclei}

As mentioned briefly earlier, the beta-decay selection rule on $K$ has been
demonstrated in charge-exchange reactions on $^{23}$Na and $^{25}$Mg
\cite{Fuji11}, but is of course also well known from heavier nuclei \cite{boh69}.

The derived formalism is directly applicable to the beta decay of odd
mass nuclei where the beta daughter through nucleon emission goes to
an even-even nucleus.
The beta-delayed proton emission from $^{21}$Mg is therefore expected to be
similar to the one from the nuclei $^{23,25}$Si and $^{27,29}$S.

For beta-delayed neutron emission the formalism is suited for initial
nuclei with odd $Z$ and even $N$, such as $^{29}$Na.

For beta-delayed $\alpha$-particle emission other cases are the nuclei $^{20}$Na and $^{22}$Al.

It could be interesting to expand the treatment to decays to odd
nuclei/odd-odd nuclei in the sd-shell, such as the decays of
$^{22}$Al, $^{24}$Si and $^{26}$P. There is as yet little relevant data
available in higher mass regions.

\section{Summary and conclusions}

Rotational bands in deformed nuclei are typically identified and
studied through the internal gamma transitions in the band. As the
excitation energy in a nucleus increases, particle emission will
eventually dominate over gamma emission and other probes of rotational
structure must be used.  This paper was motivated by a study of
beta-delayed particle emission in the lower sd-shell, i.e.\ in
relatively light nuclei away from the line of beta-stability where the
level density at excitation energies of 5--10 MeV is not too high, and
level widths are in the keV range and already dominated by particle
decays.  These states may also be populated in particle elastic
scattering or transfer reactions, but the selection rules of the beta
decay process will give a sparse feeding and give clean results. 

Our derivation resulted in selection rules, in particular for the $K$
quantum number, and the finding that structural overlap is important
for the pattern of particle emissions. These results were illustrated
through data from the beta decay of $^{21}$Mg and yielded an
interpretation of the feeding of levels in $^{21}$Na; the analysis
presented here is helpful in understanding the structure of the
particle emission pattern as well as in assigning $J$ and $K$ values
to the levels.  Rotational structure have in this way been identified
in $^{21}$Na up to the 5--10 MeV energy range.
We note that the fact that $^{21}$Mg has spin-parity $5/2^+$, but only
$K=1/2$, gives more structure information from the decay.

The simplest results depended on the assumption that it is the same
core rotating throughout the decay.  If this is not the case, the Clebsch-Gordan
selection rules will be the same, but the overlap integral (giving the
``fingerprint'' interpretation in the related case of transfer
reactions) will have a higher dimensionality and invoke the
coordinates of more particles and the interpretation correspondingly
becomes more complex.
In slightly more detail, our simple result hinges on the approximation that the
structure of the initial and final state is the same apart from the
extra particle that is emitted. In essence, it should apply
when structure development is gradual.

Apart from the clear link between beta-delayed particle emission and
transfer experiments, there is also a close correspondance to the
outgoing step in (in)elastic scattering experiments. Much of our
analysis of how the decay pattern from an excited level can be used to
extract structure information can of course be taken over (we did
not find such information in the p+$^{20}$Ne case).
Our results for beta-delayed proton emission could in part be applied also to
proton radioactivity from deformed nuclei where, however, more
accurate and detalied approaches have been employed, e.g.\ for the
decay of $^{131}$Eu in \cite{Kru00}.

In conclusion, beta-delayed particle decays just a few steps away from
the line of beta-stability can, via the particle emission process, give
access to spectrospic information that complements and extends what is
achievable via classical reaction experiments. Transfer
reactions are of course now also possible with radioactive beams, but
will typically require significantly larger intensities than beta-decay
experiments. 
Our results will be straightforward to extend and apply to higher mass regions.

\begin{acknowledgments}
  This work has been partially supported by the Independent Research
  Fund Denmark (9040-00076B).  We wish to acknowledge several illuminating
  discussions with Thomas D{\o}ssing.
\end{acknowledgments}

\appendix
  
\section{Rotations: Definitions and Properties}  \label{appA}

The notation and definitions are originally from \cite{boh69}, and
following \cite{vil66} for completeness reformulated to be directly
applicable in the present connection in \cite{sie87}.

The rotation operator, ${\cal R}_{\vec{\omega}}$ is defined by
\begin{eqnarray} \label{eqA30}
  {\cal R}_{\vec{\omega}} = \exp{(-i\phi J_{z})} \exp{(-i\theta J_{y})}
  \exp{(-i\psi J_{z})}  \; .
\end{eqnarray}
in terms of the three Euler angles, $\vec{\omega} =
(\phi,\theta,\psi)$, and the components of the angular momentum
operator, $\vec{J}=(J_x,J_y,J_z)$.  This operator describes a general
rotation from one given direction to another. Any coordinate,
$\bm{r_i}$, is then transformed into $\bm{r_i'}={\cal
  R}_{\vec{\omega}}\bm{r_i}$.

The convenient basis is $|JM\rangle$, that is the eigenstates of
$J^2$ and $J_z$.  The matrix elements of ${\cal R}_{\vec{\omega}}$
define the $D$-functions
\begin{eqnarray} \label{eqA40}
 \langle JM| {\cal R}_{\vec{\omega}} | J'M' \rangle = \delta_{J,J'}
 {\cal D}^{J*}_{MM'}(\vec{\omega}) \;.
\end{eqnarray}
The angles, $(\theta,\phi)$, are precisely defined as the angles in the
spherical harmonics, $Y_{lm}(\theta,\phi)$, and $\psi$ is the angle
describing rotation about the $z$-axis of the possibly extended
system. The relation is
\begin{eqnarray} \label{eqA50}
 {\cal D}^{J}_{M0}(\vec{\omega}) = \sqrt{4\pi/(2J+1)} Y_{JM}(\theta,\phi)\;,
\end{eqnarray}
and the orthonormality conditions are 
\begin{eqnarray} \label{eqA60}
 && \int d^3\vec{\omega} {\cal D}^{J*}_{MK}(\vec{\omega})
   {\cal D}^{J'}_{M'K'}(\vec{\omega}) \equiv \int_{0}^{\pi} \sin\theta d\theta
   \int_{0}^{2\pi} d\phi \\ \nonumber && \int_{0}^{2\pi}d\psi 
      {\cal D}^{J*}_{MK}(\vec{\omega}) {\cal D}^{J'}_{M'K'}(\vec{\omega}) =
  \frac{8\pi^2}{2J+1} \delta_{JJ'} \delta_{MM'} \delta_{KK'} \; .
\end{eqnarray}
Successive rotations of the same angles can be collected into one, that is
\begin{eqnarray} \nonumber
  {\cal D}^{J_1}_{M_1K_1}(\vec{\omega}) {\cal D}^{J_2}_{M_2K_2}(\vec{\omega}) =
\sum_{J=|J_1-j_2|}^{J_1+j_2} {\cal D}^{J}_{M_1+M_2,K_1+K_2}(\vec{\omega}) \\ 
  \langle J_1M_1 J_2M_2 |J, M_1+M_2 \rangle
  \langle J_1K_1J_2K_2 |J,K_1+K_2 \rangle. \label{eqA70} 
\end{eqnarray}

Combining successive different rotations
\begin{eqnarray} \label{eqA80}
  {\cal R}_{\vec{\omega}} &=& {\cal R}_{\vec{\omega}''} {\cal R}_{\vec{\omega}'} \\
 \nonumber  {\cal R}_{\vec{\omega}}^{-1} &=& {\cal R}_{-\vec{\omega}} =
  {\cal R}_{-\vec{\omega}'} {\cal R}_{-\vec{\omega}''} \; 
\end{eqnarray}
leads to corresponding relations between the $D$-functions
\begin{eqnarray} \nonumber
  && {\cal D}^{J*}_{MK}(\vec{\omega}) = (-1)^{M-K}{\cal D}^{J}_{-M-K}(-\vec{\omega})
 \\  &=& (-1)^{M-K} \sum_{L} {\cal D}^{J}_{-M-L}(-\vec{\omega'})
     {\cal D}^{J}_{-L-K}(-\vec{\omega''})  \label{eqA90}  \\ \nonumber
     &=& \sum_{L} {\cal D}^{J*}_{ML}(\vec{\omega'})
    {\cal D}^{J*}_{LK}(\vec{\omega''}) \; .
\end{eqnarray}

The effects on the coordinates are
\begin{eqnarray} \label{eqA100} 
  \bm{r_i'} &=&{\cal R}_{\vec{\omega}'}\bm{r_i} \;  \\ \nonumber
  {\cal R}_{\vec{\omega}}\bm{r_i} &=& 
   {\cal R}_{\vec{\omega}} {\cal R}_{\vec{\omega}'}^{-1} \bm{r_i'}
  \equiv {\cal R}_{\vec{\omega''}}\bm{r_i'} \; .
\end{eqnarray}
Transformation of a coordinate (or spin dependent) tensor,
$T_{\lambda\mu}({\bm{r_i}})$, of rank $\lambda$ is
\begin{eqnarray} \label{eqA110}  
 T_{\lambda\mu}({\bm{r_i}}) &=& T_{\lambda\mu}({{\cal R}_{\vec{\omega'}}^{-1}\bm{r_i'}})
 = \sum_{\mu'} {\cal D}^{\lambda}_{\mu \mu'}(\vec{\omega'}) 
 T_{\lambda\mu'}({\bm{r_i'}}) .
\end{eqnarray}

Finally, we shall need the following 
\begin{eqnarray} \nonumber
  \int d^3\vec{\omega} {\cal D}^{j*}_{mm'}(\vec{\omega})
  {\cal D}^{J*}_{ML}(\vec{\omega}) {\cal D}^{J'}_{M'K'}(\vec{\omega}) \\  =
  \frac{8\pi^2}{2J'+1}\langle JMjm|J'M' \rangle \langle JLjm'|J'K' \rangle  \; .
 \label{eqA120}  
\end{eqnarray}

\section{The internal wavefunction} \label{appB}
Depending on the symmetries of the internal wavefunction, the
statements in section II may have to be modified. We shall assume
here that axial symmetry is present.
Consider first the case of a nucleus with intrinsic
spin $0^+$ and $K=0$ in a rotational state $J$ (we also assume
positive parity for the intrinsic wave function). Following the derivation in
\cite{sie87}, the normalization integral in section II.C will then
apart from the contribution around $\theta = 0$ also get a
contribution around $\theta = \pi$ which will have the same value as at
$\theta = 0$, but with a phase factor $(-1)^J$. This doubles the value
of the integral for even $J$ and shows the well-known result that odd
$J$ do not occur. Other axially symmetric cases will also go from one
to two contributions, but the phase factor will become $(-1)^{J+K}$
as shown in \cite{boh69} chapter 4-2 and appendix 4A and for a
non-zero value of $K$ the projection on the internal axis of course
changes sign for the contributions at $\theta = \pi$. The case of odd
$A$ (a fermion) is slightly more complex and one must use instead the
time-reversed wavefunction $\varphi_{\bar{K'}} = - \varphi_{-K'}$.

These results must be inserted in Eq.(\ref{eq140}). The change in
the overlap integral $ME/\sqrt{N_i N_f}$ will be a doubling to two
terms (the second term with the above combined phase factor) and the
change in normalization will simply be a factor $1/\sqrt{2}$.
The expression therefore becomes
\begin{eqnarray}  \nonumber
  && \frac{ME}{\sqrt{N_i N_f}} = 4 \pi \sqrt{\frac{2J+1}{2(2J'+1)}}
  \sum_{j,l,m_l,m_l',\Sigma_p'} (-i)^l Y_{l,m_l}({\hat{\bm{k}}_p}) \\ \nonumber
  && \langle JMjm_l+\Sigma_p|J'M' \rangle
  \langle lm_ls\Sigma_p |jm_l+\Sigma_p \rangle \\ \nonumber
 &&  \left[ \langle JKjm_l'+\Sigma_p'|J'K' \rangle 
    \langle lm_l's\Sigma_p'|jm_l'+\Sigma_p' \rangle \right. \\ \label{eqB110} 
 && \int d^3 \bm{r_p} \chi^{\dagger}_{\Sigma_p'} j^{*}_{l}(k_pr_p)
  Y^{*}_{l,m_l'}({\hat{\bm{r}}_p}) \varphi_{K'}(\bm{r}_p)  - \\ \nonumber
 &&  (-1)^{J'+K'} \langle JKjm_l'+\Sigma_p'|J'-K' \rangle
   \langle lm_l's\Sigma_p'|jm_l'+\Sigma_p' \rangle \\ \nonumber
 && \left. \int d^3 \bm{r_p} \chi^{\dagger}_{\Sigma_p'} j^{*}_{l}(k_pr_p)
    Y^{*}_{l,m_l'}({\hat{\bm{r}}_p}) \varphi_{-K'}(\bm{r}_p)  \right]
    \;  . \nonumber
\end{eqnarray}

We can now in the second term change the signs of the sum indices for
$m_l'$ and $\Sigma_p'$ and use that $\langle j_1 -m_1 j_2 -m_2 | j_3
-m_3 \rangle = (-1)^{j_1+j_2-j_3} \langle j_1 m_1 j_2 m_2 | j_3 m_3 \rangle$
to simplify the expression for $K=0$ to
\begin{eqnarray}  \nonumber
  && \frac{ME}{\sqrt{N_i N_f}} = 4 \pi \sqrt{\frac{2J+1}{2(2J'+1)}}
  \sum_{j,l,m_l,m_l',\Sigma_p'} (-i)^l Y_{l,m_l}({\hat{\bm{k}}_p}) \\ \nonumber
  && \langle JMjm_l+\Sigma_p|J'M' \rangle
  \langle lm_ls\Sigma_p |jm_l+\Sigma_p \rangle \\ \nonumber
 &&  \langle J0jm_l'+\Sigma_p'|J'K' \rangle
   \langle lm_l's\Sigma_p'|jm_l'+\Sigma_p' \rangle \\ \label{eqB120} 
 && \left[ \int d^3 \bm{r_p} \chi^{\dagger}_{\Sigma_p'} j^{*}_{l}(k_pr_p)
  Y^{*}_{l,m_l'}({\hat{\bm{r}}_p}) \varphi_{K'}(\bm{r}_p)  + \right. \\ \nonumber
 && \left. (-1)^{\Xi}\int d^3 \bm{r_p} \chi^{\dagger}_{-\Sigma_p'} j^{*}_{l}(k_pr_p)
    Y^{*}_{l,-m_l'}({\hat{\bm{r}}_p}) \varphi_{-K'}(\bm{r}_p)  \right]
    \;  , \nonumber
\end{eqnarray}
where the final total phase factor becomes $\Xi = -1 + J'+K'+ (J+j-J')+ (l+s-j) 
= J+K'+l+s-1$.

For the case in section III with $K'=1/2$, $s=1/2$ and $l=0$ we end up
with a phase factor $(-1)^J$ and the two terms therefore give the same
contribution for $J$ even.
Inserting also the
numerical value of the last Clebsch-Gordan coefficient that for $K=0$ is
$\sqrt{1/2}\sqrt{(2J'+1)/(2J+1)}$ 
reduces the final result to
\begin{eqnarray}  \nonumber
  && \frac{ME}{\sqrt{N_i N_f}} = \sqrt{4 \pi}
     \langle JM\frac{1}{2}\Sigma_p|J'M' \rangle \\ \label{eq140b} 
  && \int d^3 \bm{r_p}      j^{*}_{0}(k_pr_p) \varphi_{K'}(r_p)  \; ,
\end{eqnarray}
i.e.\ the same as Eq.(\ref{eq200a}), as expected.

\end{document}